\documentclass[10pt, conference, letterpaper]{IEEEtran}
\usepackage{tikz}
\usepackage{amsthm}
\usepackage{amssymb}
\usepackage{url}
\usepackage[vlined, linesnumbered]{algorithm2e}
\SetKwIF{If}{ElseIf}{Else}{if}{}{else if}{else}{}
\SetKwFor{For}{for all}{}{}
\SetKwProg{Fn}{Function}{}{}
\DontPrintSemicolon

\newcommand\DP{\mathit{d}} 
\newcommand\SP{\mathit{s}} 
\newcommand\NH{\mathit{nh}} 
\newcommand\EQ{\mathcal{\sim}}
\newcommand\conflicts{\mathop{\#}}
\newcommand*{\quotient}[2]{\ensuremath{#1/\!\raisebox{-.65ex}{\ensuremath{#2}}}}
\newcommand\st{\,|\,}

\newtheorem{theorem}{Theorem}

\theoremstyle{definition}

\begin{document}

\title{Source-specific routing}
\author{
\IEEEauthorblockN{Matthieu Boutier and Juliusz Chroboczek}
\IEEEauthorblockA{Univ Paris Diderot, Laboratoire PPS\\
\footnotesize{Sorbonne Paris Cit\'e, PPS, UMR 7126, CNRS, F-75205 Paris, France}}
}
\date{July 2014}

\maketitle

\begin{abstract}
  Source-specific routing (not to be confused with source routing) is
  a routing technique where routing decisions depend on both the source
  and the destination address of a packet. Source-specific routing solves
  some difficult problems related to multihoming, notably in edge
  networks, and is therefore a useful addition to the multihoming
  toolbox. In this paper, we describe the semantics of source-specific
  packet forwarding, and describe the design and implementation of
  a source-specific extension to the Babel routing protocol as
  well as its implementation --- to our knowledge, the first complete
  implementation of a source-specific dynamic routing protocol ---,
  including a disambiguation algorithm that makes our implementation work
  over widely available networking APIs. We further discuss
  interoperability between ordinary next-hop and source-specific dynamic
  routing protocols. Our implementation has seen a moderate amount of
  deployment, notably as a testbed for the IETF Homenet working group.
\end{abstract}

\section{Introduction}
The routing paradigm deployed on the Internet is next-hop routing.
In next-hop routing, per-packet forwarding decisions are performed by
examining a packet's destination address only, and mapping it to
a next-hop router.  Next-hop routing is a simple, well understood paradigm
that works satisfactorily in a large number of cases.

The use of next-hop routing restricts the flexibility of the routing
system in two ways.  First, since a router only controls the next hop, a
route $A\cdot B\cdot C\cdots Z$ can only be selected by the router $A$ if
its suffix $B\cdot C\cdots Z$ has already been selected by a neighbouring
router $B$, which makes some forms of optimisation difficult or
impossible.  Other routing paradigms, such as circuit switching, label
switching and source routing, do not have this limitation.
(Source routing, in particular, has been proposed multiple times as a
suitable routing paradigm for the
Internet~\cite{saltzer1980source}, but has been discouraged due to
claimed security reasons~\cite{rfc5095}).

Second, the only decision criterion used by a router is the destination
address: two packets with the same destination are always routed in the
same manner.  Yet, there are other data in the IP header that can
reasonably be used for making a routing decision --- the TOS octet, the
IPv6 flow-id, and, of course, the source address.

We call \emph{source-specific} routing the modest extension of classical
next-hop routing where the forwarding decision is allowed to take into
account the source of a packet in addition to its destination.
Source-specific routing gives a modest amount of control over routing to
the sending host, which can choose among different routes by picking
different source addresses.  The higher layers (transport or application)
are therefore able to choose a route using standard networking APIs
(collecting the host's local addresses and binding a socket to a specific
address).  Unlike source routing, however, source-specific routing
remains a hop-by-hop mechanism, and therefore leaves local forwarding
decisions firmly in the control of the routers.

Two things are needed in order to make source-specific routing practical:
a forwarding mechanism that can discriminate on both source and
destination addresses, and a dynamic routing protocol that is able to
distribute source-specific routes.  In this paper, we describe our
experiences with the design and implementation of a source-specific
extension to the Babel routing protocol~\cite{rfc6126}, including
a \emph{disambiguation algorithm} that allows implementing source-specific
routing over existing forwarding mechanisms.

\section{Applications}

The main application of source-specific routing is the implementation of
\emph{multihoming}.

\subsection{Classical multihoming}

A multihomed network is one that is connected to the Internet through two
or more physical links.  This is usually done in order to improve
a network's fault tolerance, but can also be done in order to improve
throughput or reduce cost.

Classically, multihoming is performed by assigning
\emph{Pro\-vider-Inde\-pen\-dent} addresses to the multihomed network and
announcing them globally (in the \emph{Default-Free Zone} (DFZ)) over the
routing protocol.  The dynamic nature of the routing protocol
automatically provides for fault-tolerance; improvements in throughput and
reductions in cost can be achieved by careful engineering of the routing
protocol.

While classical multihoming works reasonably well in the network core, it
does not apply to the edge.  In order to perform classical multihoming,
a network needs to be allocated a ``Provider-Independent'' prefix that is
reannounced by some or all of a network's peers.  This setup is usually
impossible to achieve for home and small business networks.

Note that it is not in general possible to implement classical multihoming
using a single ``Provider-Dependent'' prefix.  If a network is connected
to two providers $A$ and $B$, a packet with a source address in an address
range allocated to $A$ will usually not be accepted by $B$, which will
treat it as a packet with a spoofed source address and discard it
\cite{rfc2827}.  What is more, $A$'s prefix will not be reannounced by
$B$, and hence destinations in $A$'s prefix will not be reachable over the
link to $B$.

There is some concern that classical multihoming, even when restricted to
the large networks of the core, is causing uncontrolled growth of the
``default-free routing table''.  Since we have only experimented with
source-specific routing in edge networks, we hold no opinion on the
usefulness of our techniques in the network core, and in particular on the
desirability of adding it to the BGP external routing protocol.

\subsection{Multihoming with multiple source addresses}

Since announcing the same Provider-Dependent (PD) prefix to multiple ISPs
is not always possible, it is a natural proposition to announce multiple
PD prefixes, one per provider.  In this approach, every host is assigned
multiple addresses, one per provider, and extra mechanisms are needed
(i)~to choose a suitable source and destination address for each packet,
and (ii)~to properly route each outgoing packet according to both its
source and its destination.  In a sense, using multiple addresses splits
the difficult problem of multihoming into two simpler problems that are
handled at different layers of the network stack.

\subsubsection{Choosing addresses}

The choice of source and destination addresses is typically left to the
application layer.  All destination addresses are stored within the DNS
(or explicitly carried by the application protocol), and the sending host
tries them all, either in turn \cite{rfc3484} or in parallel
\cite{rfc6555}; similarly, all possible source addresses are tried in
turn.  Once a flow is established, it is no longer possible to change the
source and destination addresses --- from the user's point of view, all
TCP connections are broken whenever a link outage forces a change of
address.  Address selection can be implemented in the operating system's
kernel and libraries, or by the application itself, which is notably done
by most modern web browsers.

A different approach is to use a transport layer that has built-in support
for multiple addresses and for dynamically renegotiating the set of source
and destination addresses.  One such transport layer is MPTCP
\cite{mptcp2012}; we describe our experience with MPTCP in
Section~\ref{sec:mptcp}.

\subsubsection{Source-specific routing}

As mentioned above, a provider will discard packets with a source address
that is in a different provider's prefix.  In a network that is connected
to multiple providers, each outgoing packet must therefore be routed
through the link corresponding to its source address.

\begin{figure}[ht]
\begin{center}
\begin{tikzpicture}[-, auto, node distance=3cm, thick, main
  node/.style={circle,draw,minimum width=.5cm}]

  \draw (-0.4, .5) node (0) {ISP 1};
  \draw (1, .5) node[main node] (1) {};
  \draw (2.5, 1) node[main node] (2) {};
  \draw (3, 0) node[main node] (3) {};
  \draw (4, 1) node[main node] (4) {};
  \draw (4.5, 0) node[main node] (5) {};
  \draw (6, .5) node[main node] (6) {};
  \draw (7.4, .5) node (7) {ISP 2};
  \draw[dashed] (3.5, .5) ellipse(2 and 1);

  \path
    (0) edge node {} (1)
    (1) edge node {} (2)
    (2) edge node {} (3)
    (2) edge node {} (4)
    (3) edge node {} (4)
    (3) edge node {} (5)
    (4) edge node {} (5)
    (5) edge node {} (6)
    (6) edge node {} (7)
    ;
\end{tikzpicture}
\end{center}
\caption{A network connected to two providers}\label{fig:two-providers}
\end{figure}
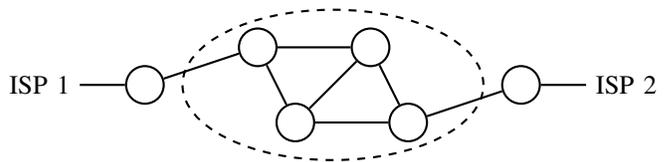

When the outgoing links are all connected to a single router, it is
feasible to set up traffic engineering rules to ensure that this happens.
There can be good reasons, however, why it is desirable to connect each
provider to a different router (Figure~\ref{fig:two-providers}): avoiding
a single point of failure, load balancing, or simply that the various
links use different link technologies that are not available in a single
piece of hardware.  In a home networking environment, the edge routers
might be provided by the different service providers, with no possibility
to consolidate their functionality in a single device.

With multiple edge routers, it is necessary that the routing protocol
itself be able to route according to source addresses.  We say that
a routing protocol performs \emph{source-specific} routing when it is able
to take both source and destination addresses into account in its routing
decisions.

\subsection{Other applications}

In addition to multihoming with multiple addresses, we are aware of two
problematic networking problems that source-specific routing solves
cleanly and elegantly.

\subsubsection{Overlay networks}

Tunnels and VPNs are commonly used to establish a net\-work-layer topology that
is different from the physical topology, notably for security reasons.  In many
tunnel or VPN deployments, the end network uses its native default route, and
only routes some set of prefixes through the tunnel or VPN.

In some deployments, however, the default route points at the tunnel.  If this
is done naively, the network stack attempts to route the encapsulated packets
through the tunnel itself, which causes the tunnel to break.  Many workarounds
are possible, the simplest being to point a host route towards the tunnel
endpoint through the native interface.

Source-specific routing provides a clean solution to that problem.  The native
default route is kept unchanged, while a source-specific default route is
installed through the tunnel.  The source-specific route being more specific
than the native default route, packets from the user network are routed through
the tunnel, while the encapsulated packets sourced at the edge router follow the
native, non-specific route.

\subsubsection{Controlled anycast}

\emph{Anycast} is a technique by which a single destination address is
used to represent multiple network endpoints.  A packet destined to an
anycast address is routed to whichever endpoint is nearest to the source
according to the routing protocol's metric.  Anycast is useful for load
balancing --- for example, the DNS root servers are each multiple
physical servers, represented by a single anycast address.

For most applications of anycast, all of the endpoints are equivalent and
it does not matter which endpoint is accessed by a given client.  Some
applications, however, require that a given user population access
a well-defined endpoint --- for example, in a Content Distribution Network
(CDN), a provider might not want to serve nodes that are not its
customers.  Ensuring that this is the case by tweaking the routing
protocol's metric (or ``prepending'' in BGP parlance) is fragile and
error-prone.

Source-specific routing provides an elegant solution to this problem.
With source-specific routing, each instance of the distributed server is
announced using a source-specific route, and will therefore only receive
packets from a given network prefix.

\section{Related work}

Multihoming is a difficult problem, and, unsurprisingly, there are many
techniques available to implement it, none of which are fully general.  In
addition to classical network-layer multihoming, already mentioned above,
there are a number of lower-layer techniques, the use of which is usually
completely transparent to the network layer; we are aware of
\emph{Multi-Link PPP}, of \emph{Ethernet link aggregation (port trunking)},
of the use of MPLS to provide multiple paths
across a rich link layer, as well as of proprietary techniques used by
vendors of cable modems.  Since these techniques work at the link
layer, they are usually restricted to multihoming with a single provider.

All of these techniques are compatible, in the sense that they can be used
at the same time.  We imagine a home network where source-specific
routing is used to access two providers, each of which is classically
multihomed, over links that consist of multiple physical links combined at
the link layer.

Source-specific packet forwarding itself is not a new idea \cite{rfc3704},
and implementing it manually on a single router using traffic engineering
interfaces is a well-documented technique \cite{lartc}.  Implementing
source-specific routing within the routing protocol has been proposed by
Bagnulo et al.\ \cite{case-sadr}, but the techniques used differ
significantly from ours.  First, the authors only deal with the
non-overlapping case --- where the different possible sources are
disjoint ---, which avoids the need for the disambiguation algorithm which
is one of our main concerns.  Second, they use a more general facility of
an existing routing protocol (BGP Communities) rather than explicitly
implementing source-specific routing.  We find our more direct approach to
be more intuitive, and expect it to be more reliable, since it doesn't
require out-of-band agreement on the meaning of the labels carried by the
routing protocol.

More generally, there are other applications of routing based on more
information from the packet header than just the destination address.  The
traffic-engineering community has been experimenting with routing based on
the TOS octet of the IPv4 header for many years, and the ability to do
that is part of the OSPFv2 protocol.  TOS-based routing is somewhat
analoguous to source-specific routing, and many of the issues raised are
similar; both can be seen as particular cases of ``multi-dimensional
routing''.

\emph{Equal Cost Multipath} (ECMP) is somewhat different.  A router
performing ECMP has multiple routes to the same destination, and chooses
among them according to the value of a hash of the packet header.  While
ECMP does route on multiple header fields, the choice of fields used to
choose a route in ECMP is a purely local matter, and does not need to be
carried by the routing protocol.

\section{Source-specific routing} \label{sec:ss-routing}

\subsection{Next-hop routing tables}

Ordinary next-hop routing consists in mapping a destination address to
a next-hop.  Obviously, it is not practical to maintain a mapping for each
possible destination address, so the mapping table must be compressed in
some manner.  The standard compressed data structure is the \emph{routing
  table} (or \emph{Forwarding Information Base}, FIB), which ranges over
\emph{prefixes}, ranges of addresses the size of which is a power of two.
The routing table can be constructed manually, but is usually populated by
a routing protocol.

Since prefixes can overlap, the routing table is an ambiguous data
structure: a packet's destination address can match multiple routing
entries.  This ambiguity is resolved by the so-called \emph{longest-prefix
  rule}: when multiple routing table entries match a given destination
address, the most specific matching entry is the one that is used.

More precisely, a prefix is a pair $P=p/\mathit{plen}$, where $p$ is the
first address in the prefix and $\mathit{plen}$ is the \emph{prefix
  length}.  An address $a$ is in $P$ when the first $\mathit{plen}$ bits
of $a$ match the first $\mathit{plen}$ bits of $p$.  We say that a prefix
$P=p/\mathit{plen}$ is more specific than a prefix $P'=p'/\mathit{plen}'$,
written $P\le P'$, when the set of addresses in $P$ is included in the set
of addresses in $P'$.  Clearly, $P\le P'$ if and only if
$\mathit{plen}\ge\mathit{plen}'$, and the first $\mathit{plen'}$ bits of
$p$ and $p'$ match.

The specificity ordering defined above has an important property: given two
prefixes $P$ and $P'$, they are either disjoint ($P \cap P' = \emptyset$), or
one is more specific than the other ($P\le P'$ or $P'\le P$).

A routing table is a set of pairs $(P, \NH)$, where $P$ is a prefix and
$\NH$, the \emph{next hop}, is a pair of an interface and a (link-local)
address; we further require that all the prefixes in a routing table be
distinct.  Because of the particular structure of prefixes, given an
address $a$, either the set of prefixes in the routing table containing
$a$ is empty, or it is a chain (a totally ordered set); hence, there
exists a most specific prefix $P$ in the routing table containing $a$.
The longest-prefix rule specifies that the next hop chosen for routing a
packet with destination $a$ is the one corresponding to this most specific
prefix, if any.

\subsection{Source-specific routing tables} \label{sec:ss-routing-tables}

Source-specific routing is an extension to next-hop routing where both the
destination and the source of a packet can be used to perform a routing
decision.  Source-specific routers use a \emph{source-specific} routing
table, which is a set of triples $(d, s, \NH)$, where $d$ is a destination
prefix, $s$ a source prefix, and $\NH$ is a next hop (note the
ordering --- destination comes first).  Such an entry matches a packet
with destination address $a_d$ and source address $a_s$ if $a_d$ is in $d$ and
$a_s$ is in $s$.
The specificity ordering generalises easily to pairs: a pair of prefixes
$(d,s)$ is more specific than a pair $(d',s')$ when all pairs of addresses
$(a_d,a_s)$ which are in $(d,s)$ are also in $(d',s')$; clearly,
$(d,s)\le(d',s')$ when $d\le d'$ and $s\le s'$.

Unfortunately, the set of destination-source pairs of prefixes equipped with
the specificity ordering does not have the same structure as the set of single
prefixes: given a pair of addresses $(a_d,a_s)$, the set of pairs of prefixes
containing $(a_d,a_s)$ might not be a chain. Consider the pairs
$(2001{:}\mathrm{db8}{:}1{::}/48,{::}/0)$ and $({::}/0,
2001{:}\mathrm{db8}{:}2{::}/48)$.  Clearly, these two pairs are not
disjoint (the pair of addresses $(2001{:}\mathrm{db8}{:}1{::}1,
2001{:}\mathrm{db8}{:}2{::}1)$ is matched by both), but neither is one
more specific than the other --- the pair $(2001{:}\mathrm{db8}{:}1{::}1,
2001{:}\mathrm{db8}{:}3{::}1)$ is matched by the first but not the second,
and, symmetrically, the pair $(2001{:}\mathrm{db8}{:}4{::}1,
2001{:}\mathrm{db8}{:}2{::}1)$ is matched by just the second.  From
a practical point of view, this means that a source-specific routing
table can contain multiple most-specific entries, and thus fail to
unambiguously specify a forwarding behaviour.

We say that a source-specific routing table is \emph{ambiguous} when it
contains multiple non-disjoint most-specific entries.  Two entries $r_1$
and $r_2$ that are neither disjoint nor ordered are said to be
\emph{conflicting}, written $r_1\conflicts r_2$.  If $r_1=(d_1,s_1)$ and
$r_2=(d_2,s_2)$, then this is equivalent to saying that either $d_1<d_2$
and $s_1>s_2$ or $d_1>d_2$ and $s_1<s_2$.  We call the \emph{conflict
  zone} of $r_1$ and $r_2$ the set of $(a_d,a_s)$ that are matched by both
$r_1$ and $r_2$.

\subsection{Forwarding behaviour}

In the presence of an ambiguous routing table, there exist packets that
are matched by distinct most-specific entries.  An arbitrary choice must
be made in order to decide how to route such a packet.

\begin{figure}[ht]
\begin{center}
\begin{tikzpicture}[-, auto, node distance=3cm, thick, main
  node/.style={circle,draw}]

  \draw (-0.4, 0) node (0) {};
  \draw (1, 0) node[main node] (1) {A};
  \draw[->] (0.6, 0.2) -- (0.1,0.2) node[above] {$(d_1,s_1)$};
  \draw[->] (1.4, -0.2) -- (1.9,-0.2) node[below] {$(d_2,s_2)$};

  \draw (4, 0) node[main node] (2) {B};
  \draw[->] (3.6, 0.2) -- (3.1,0.2) node[above] {$(d_1,s_1)$};
  \draw[->] (4.4, -0.2) -- (4.9,-0.2) node[below] {$(d_2,s_2)$};
  \draw (5.2, 0) node (3) {};

  \path
    (0) edge node {} (1)
    (1) edge node {} (2)
    (2) edge node {} (3)
    ;
\end{tikzpicture}
\end{center}
\caption{A routing loop due to incoherent orderings}\label{fig:routing-loop}
\end{figure}
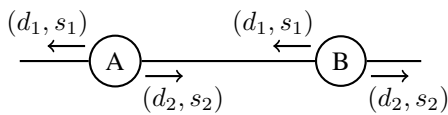

Let us first remark that all routers in a single routing domain must make
a consistent choice --- having different routers follow different policies
within conflict zones may lead to persistent routing loops.  Consider the
topology in Figure~\ref{fig:routing-loop}, with two source-specific routes
indexed by the pairs $(d_1, s_1)$ and $(d_2, s_2)$ respectively, where
packets matching $(d_1,s_1)$ are sent towards the left of the diagram, and
packets matching $(d_2,s_2)$ are sent towards the right.  If the two
pairs are in conflict, and router $A$ chooses $(d_2,s_2)$ while $B$
chooses $(d_1,s_1)$, then a packet matching both pairs will loop between
$A$ and $B$ indefinitely.

It is therefore necessary to choose a disambiguation rule that is uniform
across the routing domain.  There are two natural choices: discriminating
on the destination first, and comparing sources if destinations are equal,
or discriminating on source first.  More precisely, the destination-first
ordering is defined by:
\[ (d,s)\preceq (d',s')\mbox{ if }d < d'\mbox{ or }d=d'\mbox{ and }s\le s',\]
while the source-first ordering is defined by
\[ (d,s)\preceq_s (d',s')\mbox{ if }s<s'\mbox{ or }s=s'\mbox{ and }d\le d'.\]
These orderings are isomorphic --- hence, there is no theoretical argument
that allows us to choose between them.  An engineering choice must be
made, based on usefulness alone.

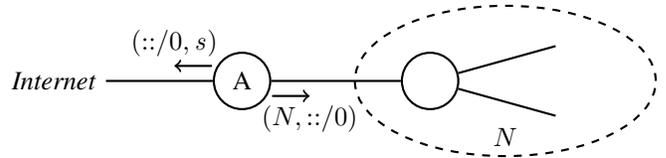
\begin{figure}[ht]
\begin{center}
\begin{tikzpicture}[-, auto, node distance=3cm, thick, main
  node/.style={circle,draw,minimum width=.75cm}]

  \draw (-1, 0) node (0) {\emph{Internet}};
  \draw (1.5, 0) node[main node] (1) {A};
  \draw[->] (0.6 + .5, 0.2) -- (0.1 + .5,0.2) node[above] {$({::}/0, \SP)$};
  \draw[->] (1.4 + .5, -0.2) -- (1.9 + .5,-0.2) node[below] {$(N, {::}/0)$};
  \draw (4, 0) node[main node] (2) {};
  \draw (5.8, .5) node (3) {};
  \draw (5.8, -.5) node (4) {};
  \draw[dashed] (5, 0) ellipse(2 and 1);
  \draw (5, -1) node[above] {$N$};

  \path
    (0) edge node {} (1)
    (1) edge node {} (2)
    (2) edge node {} (3)
    (2) edge node {} (4)
    ;
\end{tikzpicture}
\end{center}
\caption{A stub network behind a source-specific router}
\label{fig:destination-first}
\end{figure}

The current consensus, both within the IETF Homenet group and outside it,
appears to be that the destination-first ordering is the more useful of
the two.  Consider the (fairly realistic) topology in
Figure~\ref{fig:destination-first}, where an edge router $A$ announces
a source-specific route towards the Internet, and a stub network $N$
announces a (non-specific) route to itself.  A packet matching both routes
must follow the route towards $N$, since it is obviously the only route
that can reach the destination, which implies that $A$ must use the
destination-first ordering.  On the other hand, we know of no such
compelling examples of the usefulness of the source-first ordering.

In the following sections, we describe our experience with
source-specific routing using the destination-first ordering.  However,
nothing in this article depends on the particular ordering being used,
and our techniques would apply just as well to any structure that is a
refinement of the specificity ordering and that is totally ordered on
route entries containing a given address.

\section{Implementing source-specific routing}

In the previous sections, we have described source-specific routing and
shown how all routers in a routing domain must make the same choices with
respect to ambiguous routing tables, and have argued in favour of the
destination-first semantics.  Whichever particular choice is made by an
implementation of a routing protocol, however, must be implementable in
terms of the primitives made available by the lower layers (the operating
system kernel and the hardware).

In this section, we describe the two techniques that we have used to
implement a source-specific extension to the Babel routing protocol
\cite{babel-source-specific}.  We first describe the technique that we use
when running over a lower layer that natively implements destination-first
source-specific routing (Section~\ref{sec:native-fib}).  We then describe
our so-called ``disambiguation'' algorithm
(Section~\ref{sec:disambiguating}) which we use to implement
destination-first source-specific routing over any source-specific
facility provided by the lower layers, as long as it is compatible with
the specificity ordering --- a very mild hypothesis that is satisfied by
a number of widely available implementations.

\subsection{Native source-specific FIB} \label{sec:native-fib}

Ideally, we would like the lower layers of the system (the OS kernel, the
line cards, etc.) to implement destination-first source-specific routing
tables out of the box.  Such native support for source-specific routing is
preferable to the algorithm described below, since no additional routes
will need to be installed.  In practice, while many systems have
a facility for source-specific traffic engineering, this lower-layer
support often has a behaviour different from the one that we require.

The Linux kernel, when compiled with the relevant options
(``ipv6-subtrees''), supports source-specific FIBs natively, albeit for
IPv6 only.  Unfortunately, this support is only functional since Linux
3.11 (source-specific routes were treated as unreachable in earlier
versions), and only for IPv6 (for IPv4, the ``source'' datum is silently
ignored).  We know of no other TCP/IP stacks with native support for
destination-first source-specific routing --- other techniques must be
used on most systems.

\subsection{Disambiguation of a routing table} \label{sec:disambiguating}

All versions of Linux, some versions of FreeBSD, and a number of other
networking stacks implement a facility to manipulate multiple routing
tables and to select a particular one depending on the source address of
a packet.  Since the table is selected before the destination address is
examined, these API implement the source-first behaviour, which is not
what we aim to implement.

In this section, we describe a disambiguation algorithm that can be used
to maintain a routing table that is free of ambiguities, and will
therefore yield the same behaviour as long as the underlying forwarding
mechanism implements a behaviour that is compatible with the specificity
ordering (Section~\ref{sec:ss-routing-tables}).  All the forwarding
mechanisms known to us satisfy this very mild hypothesis.

Recall that a routing table is ambiguous if there exists a packet that is
matched by at least one entry in the table and such that there is no
most-specific entry among the matching entries.  A necessary and
sufficient property for a routing table to be non-ambiguous is that every
conflict zone is equal to the union of more specific route entries.

The algorithm that we propose maintains, for each conflict, exactly one
route entry that covers exactly the conflict zone.  While a more
parsimonious solution would be possible in some cases, it would greatly
complicate the algorithm.

\paragraph{Weak completeness}
We say that a routing table is \emph{weakly complete} if each conflict zone is
covered by more specific entries.  More formally, $T$ is weakly complete if
$\forall r_1, r_2 \in T, r_1 \cap r_2 = \bigcup \{r \in T \mid r \le r_1 \cap
r_2\}$.

\begin{theorem}
A routing table is non-ambiguous if and only if it is weakly complete.
\end{theorem}

\begin{IEEEproof}
  Let $U_x^y = \bigcup \{r \in T \st r \le x \cap y\}$.  We need to show that $T$
  is non-ambiguous iff $\forall r_1, r_2 \in T, r_1 \cap r_2 = U_{r_1}^{r_2}$.

  ($\Leftarrow$) Suppose $T$ is weakly complete, and consider two route entries
  $x, y \in T$ in conflict.  By weak completeness, $U_x^y = x \cap y$, so for
  all addresses $a \in x \cap y$, there exists a route $r \in U_x^y$ such that
  $a \in U_x^y$.  Since $r \in x \cap y$, we have $r < x$ and $r < y$, and $r$
  is more specific than $x \cap y$.  Since this is true for all conflicts, the
  table is not ambiguous.

  ($\Rightarrow$) Suppose $T$ is non-ambiguous and not weakly complete.  Then
  there exist two entries $x, y \in T$ in conflict such that $x \cap y \neq
  U_x^y$.  Consider an address $a \in x \cap y \smallsetminus U_x^y$, and an
  entry $r \in T$ matching $a$.  Clearly, $r \supsetneq x \cap y$, and so either
  $r \conflicts x$ or $r \conflicts y$, or $r > x$ and $r > y$.  In all cases,
  $r$ is not more specific than both $x$ and $y$, so there is no minimum for the
  set of entries matching $a$.  This contradicts the hypothesis, so if $T$ is
  not ambiguous, it is weakly complete.
\end{IEEEproof}

Disambiguation with weak completeness is not convenient, since it may require
adding multiple route entries to solve a single conflict, and the disambiguation
routes added may generate additional conflicts.  Suppose for example that the
FIB first contains two entries $r_1 > r_2$, and we add $r_3 > r_2$ which
conflicts with $r_1$ (see figure below).  Since $r_2 < r_3$, there is no
conflict within $r_2$, but we need disambiguation routes $d_1$ and $d_2$.  The
FIB is now weakly complete.

Suppose now that we add $r_4 < r_3$ in conflict both with $r_1$ and the
disambiguation route $d_2$.  We install a new disambiguation entry $d_3$.  Note
also that since $r_4 < r_3$, we need to use the next-hop of $r_4$ for the former
region covered by $d_1$: we need to change the currently installed
disambiguation route entry.

\begin{center}
\begin{tikzpicture}[scale=0.65]
  \begin{scope}
    \draw (0,0) -- (0,4) -- (2,4) -- (2,0) -- cycle;
    \draw (0, 0) node[above right] {$r_1$};
    \draw (0,2) -- (0,3) -- (1,3) -- (1,2) -- cycle;
    \draw (1, 2) node[above left] {$r_2$};
  \end{scope}
  \draw (2.5, 2) node {$\rightarrow$};
  \begin{scope}[xshift=3cm]
    \draw (0,0) -- (0,4) -- (2,4) -- (2,0) -- cycle;
    \draw (0, 0) node[above right] {$r_1$};
    \draw (0,2) -- (0,3) -- (1,3) -- (1,2) -- cycle;
    \draw (1, 2) node[above left] {$r_2$};
    \draw[very thick] (0,2) -- (0,4) -- (4,4) -- (4,2) -- cycle;
    \draw (4,2) node[above left] {$r_3$};
    \draw [dashed]
    (0 +0.1, 3 +0.1) -- (0 +0.1, 4 -0.1) --
    (1 -0.1, 4 -0.1) -- (1 -0.1, 3 +0.1) -- cycle;
    \draw (0.5, 3.5) node {$d_1$};
    \draw [dashed]
    (1 +0.1, 2 +0.1) -- (1 +0.1, 4 -0.1) --
    (2 -0.1, 4 -0.1) -- (2 -0.1, 2 +0.1) -- cycle;
    \draw (1.5, 3) node {$d_2$};
  \end{scope}
  \draw (7.5, 2) node {$\rightarrow$};
  \begin{scope}[xshift=8cm]
    \draw (0,0) -- (0,4) -- (2,4) -- (2,0) -- cycle;
    \draw (0, 0) node[above right] {$r_1$};
    \draw (0,2) -- (0,3) -- (1,3) -- (1,2) -- cycle;
    \draw (1, 2) node[above left] {$r_2$};
    \draw (0,2) -- (0,4) -- (4,4) -- (4,2) -- cycle;
    \draw (4,2) node[above left] {$r_3$};
    \draw
    (0 +0.01, 3 +0.01) -- (0 +0.01, 4 -0.01) --
    (1 -0.01, 4 -0.01) -- (1 -0.01, 3 +0.01) -- cycle;
    \draw
    (1 +0.01, 2 +0.01) -- (1 +0.01, 4 -0.01) --
    (2 -0.01, 4 -0.01) -- (2 -0.01, 2 +0.01) -- cycle;
    \draw[very thick] (0,3) -- (0,4) -- (3,4) -- (3,3) -- cycle;
    \draw (3,3) node[above left] {$r_4$};
    \draw [dashed]
    (0 +0.1, 3 +0.1) -- (0 +0.1, 4 -0.1) --
    (1 -0.1, 4 -0.1) -- (1 -0.1, 3 +0.1) -- cycle;
    \draw (0.5, 3.5) node {$d_1$};
    \draw [dashed]
    (1 +0.1, 3 +0.1) -- (1 +0.1, 4 -0.1) --
    (2 -0.1, 4 -0.1) -- (2 -0.1, 3 +0.1) -- cycle;
    \draw (1.5, 3.5) node {$d_3$};
  \end{scope}
\end{tikzpicture}
\end{center}

Some of this complexity can be avoided by requiring a stronger notion of
completeness.

\paragraph{Completeness}
A routing table is \emph{(strongly) complete} if each conflict zone is
covered by one route entry.  More formally, $T$ is complete if $\forall
r_1, r_2 \in T, r_1 \cap r_2 \in T$.  This obviously implies
weak-completeness, and therefore a complete routing table is not
ambiguous.  Our algorithm maintains the completeness of the routing table.

\begin{theorem}
  Adding routes to achieve completeness does not lead to \emph{another}
  conflict.
\end{theorem}

\begin{IEEEproof}
  Suppose that $r_1 = (d_1, s_1)$ and $r_2 = (d_2, s_2)$ are two route entries
  in conflict, where $d_1 < d_2$ and $s_1 > s_2$.  Consider the disambiguation
  entry $r_{sol} = (d_1, s_2)$ which disambiguates this conflict.  Suppose now
  that $r_{sol}$ is in conflict with another route entry $r_3 = (d_3, s_3)$.  We
  have either $d_1 < d_3$ and $s_1 > s_2 > s_3$, in which case $r_3 \conflicts
  r_1$ ; or $d_2 > d_1 > d_3$ and $s_2 < s_3$, in which case $r_3 \conflicts
  r_2$.  In either case, the conflict existed beforehand, and must therefore
  already have been resolved.
\end{IEEEproof}

Take the previous example again.  When adding $r_3$, we add one route entry to
cover the area $d_1$ ($r_1 \cap r_3$).  Since $r_2$ is more specific, the new
route entry does not affect the routing decision for addresses in $r_2$.  When
adding $r_4$, it is in conflict with both $r_1$ and the disambiguation route
$d_1$, but for the same conflict zone $r_4 \cap r_1$.  The disambiguation
route inserted is thus not an additional conflict.

\begin{center}
\begin{tikzpicture}[scale=0.65]
  \begin{scope}
    \draw (0,0) -- (0,4) -- (2,4) -- (2,0) -- cycle;
    \draw (0, 0) node[above right] {$r_1$};
    \draw (0,2) -- (0,3) -- (1,3) -- (1,2) -- cycle;
    \draw (1, 2) node[above left] {$r_2$};
  \end{scope}
  \draw (2.5, 2) node {$\rightarrow$};
  \begin{scope}[xshift=3cm]
    \draw (0,0) -- (0,4) -- (2,4) -- (2,0) -- cycle;
    \draw (0, 0) node[above right] {$r_1$};
    \draw (0,2) -- (0,3) -- (1,3) -- (1,2) -- cycle;
    \draw (1, 2) node[above left] {$r_2$};
    \draw[very thick] (0,2) -- (0,4) -- (4,4) -- (4,2) -- cycle;
    \draw (4,2) node[above left] {$r_3$};
    \draw [dashed]
    (0 +0.1, 2 +0.1) -- (0 +0.1, 4 -0.1) --
    (2 -0.1, 4 -0.1) -- (2 -0.1, 2 +0.1) -- cycle;
    \draw (1.5, 3.5) node {$d_1$};
  \end{scope}
  \draw (7.5, 2) node {$\rightarrow$};
  \begin{scope}[xshift=8cm]
    \draw (0,0) -- (0,4) -- (2,4) -- (2,0) -- cycle;
    \draw (0, 0) node[above right] {$r_1$};
    \draw (0,2) -- (0,3) -- (1,3) -- (1,2) -- cycle;
    \draw (1, 2) node[above left] {$r_2$};
    \draw (0,2) -- (0,4) -- (4,4) -- (4,2) -- cycle;
    \draw (4,2) node[above left] {$r_3$};
    \draw
    (0 +0.01, 2 +0.01) -- (0 +0.01, 4 -0.01) --
    (2 -0.01, 4 -0.01) -- (2 -0.01, 2 +0.01) -- cycle;
    \draw (1.5, 2.5) node {$d_1$};
    \draw[very thick] (0,3) -- (0,4) -- (3,4) -- (3,3) -- cycle;
    \draw (3,3) node[above left] {$r_4$};
    \draw [dashed]
    (0 +0.1, 3 +0.1) -- (0 +0.1, 4 -0.1) --
    (2 -0.1, 4 -0.1) -- (2 -0.1, 3 +0.1) -- cycle;
    \draw (1, 3.5) node {$d_2$};
  \end{scope}
\end{tikzpicture}
\end{center}

\paragraph{Preliminaries}

We write $\min(r_1, r_2)$ for the minimum according to $\preceq$.  We
define two auxiliary functions.  The function
$\mathrm{min\_conflict}(\mathit{zone}, r)$
(Algorithm~\ref{alg:min-conflict}) returns, if it exists, the minimum
route entry in conflict with $r$ for the conflict zone $\mathit{zone}$.
The function $\mathrm{conflict\_solution}(\mathit{zone})$
(Algorithm~\ref{alg:conflict-solution}) returns, if it exists, the
minimum route entry participating in a conflict for the zone
$\mathit{zone}$.

\begin{algorithm}
\caption{search for mininum conflicting route}\label{alg:min-conflict}
\Fn({min\_conflict($\mathit{zone}, r$)}){}{
  $\mathit{min} \leftarrow \bot$\;
  \For{$r_1 \in T$\\
    s.t. $r \conflicts r_1$ and $r \cap r_1 = \mathit{zone}$}{
    $\mathit{min} \leftarrow \min(r_1, \mathit{min})$
  }
  return $\mathit{min}$
}
\end{algorithm}

\begin{algorithm}
\caption{Search for conflict solution}\label{alg:conflict-solution}
\Fn({conflict\_solution($\mathit{zone}$)}){}{
  $\mathit{min} \leftarrow \bot$\;
  \For{$r_1, r_2 \in T$\\
    s.t. $r_1 \conflicts r_2$ and $r_1 \cap r_2 = \mathit{zone}$
    and $r_1 \prec r_2$}{
    $\mathit{min} \leftarrow \min(r_1, \mathit{min})$
  }
  return $\mathit{min}$
}
\end{algorithm}

We write $\mathrm{nh}(r)$ for the next hop of a route $r$.

We use three primitives for manipulating the routing table.
Let $r=(d, s, \NH)$ be a route entry, and $\NH'$ a nexthop.  Then
$\mathrm{install}(r, \NH')$ adds the route entry $(d, s, \NH')$,
$\mathrm{uninstall}(r, \NH')$ removes the route entry $(d, s, \NH')$,
and $\mathrm{switch}(r, \NH', \NH'')$ changes the FIB's
route entry $(d, s, \NH')$ to $(d, s, \NH'')$.  Calling $\mathrm{switch}(r,
\NH', \NH'')$ is equivalent to calling $\mathrm{uninstall}(r, \NH')$
followed by $\mathrm{install}(r, \NH'')$.

\paragraph{Relevant conflicts}
Consider a route entry $r$, and a set $E$ of routing entries in conflict with
$r$ for the same conflict zone; all of these conflicts will have the same
resolution.  Moreover, if the resolution was caused by a route in $E$, then that
was necessarily the more specific of the entries in $E$.  Note that the minimum
exists because elements of $E$ have either the same destination, or the same
source, and match at least one address in $r$.

Given a route entry $r$, we define the equivalence $\EQ_r$ by $r_1 \EQ_r r_2
\Leftrightarrow r_1 \cap r = r_2 \cap r$, i.e. two route entries are equivalent
for $\EQ_r$ if they have the same intersection with $r$.  If two equivalent
route entries are in conflict with $r$, this means that they have the same
conflict zone.

Quotienting a set of routing entries in conflict with $r$ by this equivalence,
and taking the minimum of each of the class of equivalence gives us exactly the
routes that we care about.

\paragraph{Adding a route entry (Algorithm~\ref{alg:add-route})}
Installing a new route entry in the FIB may make it ambiguous.  For this
reason, we must install the most specific routing entries first.  In
particular, we must install disambiguation entries (lines 2 to 9) before
the route itself (lines 10 to 14).

Let $r$ be the route to install, and $C$ the set of route entries in
conflict with $r$, for which there is no natural solution, i.e. $C = \{r'
\in T \mid r' \conflicts r\ \mathrm{and}\ r' \cap r \not\in T\}$ (line
3).  We only consider the relevant conflicts upon this set (line 4): $C' =
\{\min(E) \mid E \in \quotient{C}{\EQ_{r}}\}$.

For each route entry $r_1 \in C'$ (considering the most specific first),
we first search (line 5), if it exists, the minimum route entry $r_2$ such
that $r_2 \conflicts r_1$ and $r_2 \cap r_1 = r \cap r_1$.  If $r_2$ does
not exist, then there was no conflict for this zone before, and we must
add $((r_1 \cap r), \NH)$ to the FIB (line 7).  Otherwise, a routing entry
has been installed for this conflict, and we must decide if the new route
entry $r$ is or not the new candidate, which is true if it is more
desirable $(\preceq)$ than both $r_2$ and $r_1$ (line 8).  If it is the
case, then the previous next-hop installed was the one of $r_2$: we
replace $((r_1 \cap r), \NH_2)$ by $((r_1 \cap r), \NH)$ (line 9).

Finally, we must search if there exists two route entries in conflict for
the zone of $r$ (line 10).  In that case, a disambiguation route entry has
been installed, so $r$ must replace it (line 12).  Otherwise, $r$ can be
added normally (line 14).  We end the procedure by adding $r$ to our local
RIB (line 15).

\begin{algorithm}
\label{alg:add-route}
\caption{Route addition}
\Fn({add\_route($r$)}){}{
  \For{$r_1 \in T$\\
    s.t. $r \conflicts r_1$ and $r \cap r_1 \not\in T$\\
    and $r_1 = \mathrm{min\_conflict}(r \cap r_1, r)$}{
    $r_2 \leftarrow \mathrm{min\_conflict}(r \cap r_1, r_1)$\;
    \If{$r_2 = \bot$}{
      $\mathrm{install}(r \cap r_1, \mathrm{nh}(\min(r, r_1)))$
    } \ElseIf{$r \prec r_2$ and $r \prec r_1$}{
      $\mathrm{switch}(r \cap r_1, \mathrm{nh}(r_2), \mathrm{nh}(r))$
    }
  }
  $r_1 \leftarrow \mathrm{conflict\_solution}(r)$\;
  \If{$r_1 = \bot$}{
    $\mathrm{install}(r, \mathrm{nh}(r))$
  } \Else {
    $\mathrm{switch}(r, \mathrm{nh}(r_1), \mathrm{nh}(r))$
  }
  $T \leftarrow T \cup \{r\}$\;
}
\end{algorithm}

\paragraph{Removing a route entry (Algorithm~\ref{alg:delete-route})}
This time, we must first remove the less specific route first to keep the
routing table unambiguous.  Again, we write $r$ for the route to be
removed.  First, remove $r$ from the RIB (line 2).  As for the addition,
$r$ may be solving a conflict, in which case we cannot just remove it,
but must first search for the entry covering that conflict (line 3), and
if it exists replace $r$'s next-hop (line 7).  Otherwise, we just remove
$r$ from the FIB (line 5).

We consider $C'$ as previously defined (lines 9 and 10).  For each route
entry $r_1 \in C'$ (considering the less specific first), we first search,
as we did for the adding process, for the minimum route entry $r_2$ such
that $r_2 \conflicts r_1$ and $r_2 \cap r_1 = r \cap r_1$ (line 11).  If
$r_2$ does not exist, we remove $((r_1 \cap r), \NH)$ from the FIB (line
13).  Otherwise, for the same reasons above, if $r$ is more desirable than
both $r_1$ and $r_2$, then we replace in the FIB the next-hop of $r$
assigned for $r \cap r_1$ by the one of $r_2$ (line 15).

\begin{algorithm}
\label{alg:delete-route}
\caption{Route deletion}
\Fn({delete\_route($r$)}){}{
  $T \leftarrow T \smallsetminus \{r\}$\;
  $r_1 \leftarrow \mathrm{conflict\_solution}(r)$\;
  \If{$r_1 = \bot$}{
    $\mathrm{uninstall}(r, \mathrm{nh}(r))$
  } \Else {
    $\mathrm{switch}(r, \mathrm{nh}(r), \mathrm{nh}(r_1))$
  }
  \For{$r_1 \in T$\\
    s.t. $r \conflicts r_1$ and $r \cap r_1 \not\in T$\\
    and $r_1 = \mathrm{min\_conflict}(r \cap r_1, r)$}{
    $r_2 \leftarrow \mathrm{min\_conflict}(r \cap r_1, r_1)$\;
    \If{$r_2 = \bot$}{
      $\mathrm{uninstall}(r \cap r_1, \mathrm{nh}(\min(r, r_1)))$
    } \ElseIf{$r \prec r_2$ and $r \prec r_1$}{
      $\mathrm{switch}(r \cap r_1, \mathrm{nh}(r), \mathrm{nh}(r_2))$
    }
  }
}
\end{algorithm}

\paragraph{Changing a route entry (Algorithm~\ref{alg:change-route})}
This is the simplest case, since disambiguation routes must be maintained,
and changed only if the route that we want to change has been selected for
disambiguation.  The order in which we change the route entries does not
matter.  Let $r$ the route entry to change by $r_{\mathit{new}}$.  Here,
we choose to first replace $r$ by $r_{\mathit{new}}$ (line 2).

We consider $C'$ as previously defined (lines 3 and 4).  For each route
entry $r_1 \in C'$, we search for the minimum route entry $r_2$ such that
$r_2 \conflicts r_1$ and $r_2 \cap r_1 = r \cap r_1$.  If both $r \prec
r_1$ and $r_2$ is $r$ (line 6), then we replace the next-hop $\NH$ of the
corresponding disambiguation route entry by the new one
$\NH_{\mathit{new}}$ (line 7).

\begin{algorithm}
\label{alg:change-route}
\caption{Route modification}
\Fn({change\_route($r, r_{\mathit{new}}$)}){}{
  $\mathrm{switch}(r, \mathrm{nh}(r), \mathrm{nh}(r_{\mathit{new}}))$\;
  \For{$r_1 \in T$\\
    s.t. $r \conflicts r_1$ and $r \cap r_1 \not\in T$\\
    and $r_1 = \mathrm{min\_conflict}(r \cap r_1, r)$\\
    and $r \prec r_1$
    and $r = \mathrm{min\_conflict}(r \cap r_1, r_1)$}{
    $\mathrm{switch}(r \cap r_1, \mathrm{nh}(r), \mathrm{nh}(r_{\mathit{new}}))$
  }
}
\end{algorithm}

\subsection{External changes to the routing table}

In the description above, we have asssumed that only our algorithm ever
needs to manipulate the routing table.  In practice, however, the routing
table is also manipulated by other agents --- other routing protocols or
human operators.  In principle, the same algorithm should be applied to
externally changed routes; however, this is not implemented yet.

\section{Source-specific Bellman-Ford} \label{sec:bf}

The distributed Bellman-Ford algorithm is the foundation of a number of
more or less widely deployed routing protocols, such as the venerable RIP,
EIGRP, Babel and, arguably, BGP.  In order to experiment with
source-specific routing, we have implemented a source-specific variant of
the Babel routing protocol~\cite{rfc6126}; the exact details of the packet
format of our extension are described in \cite{babel-source-specific}.
Our implementation has seen a moderate amount of deployment, most notably
as a testbed for the IETF Homenet working group \cite{homenet-arch}.

Ordinary (next-hop) distributed Bellman-Ford maintains a \emph{routing
  table} which associates, to each known destination prefix, a next-hop
router and a metric; each prefix and metric pair is advertised to
neighbours in periodic \emph{update} messages.  In source-specific
Bellman-Ford, the routing table is indexed by pairs of a destination
prefix and a source prefix, and (source-specific) updates advertise a
triple of a destination prefix, a source prefix, and a metric.

The source-specific extension to Babel adds a new kind of source-specific
update message in addition to the original, non-specific update.  Since
Babel's loop-avoidance mechanism relies on two kinds of request messages,
it also adds two new kinds of source-specific requests.  All of these are
encoded as new kinds of messages rather than extensions to existing
messages, which causes them to be silently ignored by unextended Babel
routers, and ensures that our extension interoperates with the
original Babel protocol.

\subsection{Bootstrapping}

In distributed Bellman-Ford, a prefix is reannounced after it has been
learnt from a neighbour.  This process is bootstrapped by announcing
prefixes learned from a different source (typically a different routing
protocol or a static route); in Babel, this is known as
\emph{redistribution}.

Just like ordinary routes, source-specific routes are originated by
performing redistribution.  In case a source-specific route is already
present, our implementation is able to redistribute it; more generally,
the filtering language allows attaching a source prefix to a non-specific
route at redistribution time.  While careless use of this facility may
cause persistent routing loops to occur, this is expected with careless
redistribution.

\subsection{Interoperability}

The Babel protocol has seen a moderate amount of deployment in production
networks, and is usually deployed within cheap routers that can be
difficult to update with a source-specific version of the protocol.  We
have therefore paid particular attention to the issue of interoperability
between routers running the source-specific and unextended protocols.

The extended version of the protocol uses both non-specific and specific
update messages.  In principle, a non-specific route could be announced in
two manners: by using a non-specific update carrying the destination
prefix $\DP$, or by using a source-specific update carrying the pair
$(\DP, {::}/0)$.  As we want non-specific routes to be propagated between
source-specific and non-specific routers, source-specific routers
interpret a non-specific update as a source-specific update with
a source prefix of ${::}/0$, and, conversely, source-specific routers never
send source-specific updates of the form $(\DP, {::}/0)$, preferring the
non-specific form instead.

A more difficult issue is how a non-specific router should interpret
a source-specific update.  There are two possibilities: the source can be
discarded and the update treated as non-specific, or the entire update can
be discarded.  The first of these possibilities can cause persistent
routing loops.

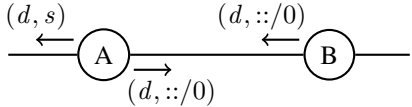
\begin{figure}[ht]
\begin{center}
\begin{tikzpicture}[-, auto, node distance=3cm, thick, main
  node/.style={circle,draw}]

  \draw (-0.4, 0) node (0) {};
  \draw (1, 0) node[main node] (1) {A};
  \draw[->] (0.6, 0.2) -- (0.1,0.2) node[above] {$(\DP, \SP)$};
  \draw[->] (1.4, -0.2) -- (1.9,-0.2) node[below] {$(\DP, {::}/0)$};

  \draw (4, 0) node[main node] (2) {B};
  \draw[->] (3.6, 0.2) -- (3.1,0.2) node[above] {$(\DP, {::}/0)$};
  \draw (5.2, 0) node (3) {};

  \path
    (0) edge node {} (1)
    (1) edge node {} (2)
    (2) edge node {} (3)
    ;
\end{tikzpicture}
\end{center}
\caption{Non-specific routers cannot accept specific routes}
\label{fig:routing-loop-2}
\end{figure}

Consider two nodes A and B, with A source-specific announcing a route to
$(\DP, \SP)$ (Figure~\ref{fig:routing-loop-2}).  Suppose that B ignores the
source information when it receives the update, and reannounces it as
$\DP$.  This is reannounced to A, which treats it as $(\DP, {::}/0)$.
Packets destined to $\DP$ but not sourced in $\SP$ will be forwarded by
A to B, and by B to A, causing a persistent routing loop.

On the other hand, if non-source-specific nodes reject source-specific
updates, but source-specific nodes accept non-specific updates, then
source-specific nodes can communicate entries of the form $(\DP, {::}/0)$
and are completely compatible with non-source-specific nodes.  In this
case, Bellman-Ford will eventually converge to a loop-free configuration.

In general, discarding of source-specific routes by non-specific routers
will cause routing blackholes.  Intuitively, unless there are enough
non-specific routes in the network, non-specific routers will suffer
starvation, and discard packets for destinations that are only announced
by source-specific routers.  A simple yet sufficient condition for
avoiding blackholes is to build a connected source-specific backbone that
includes all of the edge routers, and announce a (non-specific) default
route towards the backbone.

\section{Experimental results} \label{sec:experiment}

We have implemented both schemes described in
Sections~\ref{sec:native-fib} and~\ref{sec:disambiguating} within
\texttt{babeld}, a Linux implementation of the Babel routing protocol.
This has allowed us to perform a number of experiments which we describe
in this section.

Our experimental network consists of a mesh network consisting of a dozen
OpenWRT routers and a single server running Debian Linux.  Two of the mesh
routers have a wired connection to the Internet, and are connected to the
server through VPNs (over IPv4).  All of the routers run our modified
version of the Babel protocol.

IPv4 connectivity for the mesh is provided by the Debian server, which
acts as a NAT box.  The IPv6 connectivity is more interesting: there are
two IPv6 prefixes, one of which is a native prefix provided by our
employer's network, the other one being routed through the VPN.  The
network therefore has two source-specific default IPv6 routes.

\subsection{Routing table for VPN connectivity}
Figure~\ref{fig:routing-table} shows an excerpt of the routing tables of one of
the two wired routers.  The modified \texttt{babeld} daemon has allocated a
non-default routing table, table 11, and inserted routes (marked as
\texttt{proto 42}) into both the default \texttt{main} table and table 11.  The
former contains non-specific routes: the default route and the
\texttt{/20} subnet announced by our local DHCP server, and host routes
to individual mesh nodes.  The encapsulated VPN packets are routed through
the default route.

Table 11 contains routes for locally originated packets, sourced in
192.168.4.0/24.  The only ``real'' route in this table is the default route,
which prevents the VPN from attempting to ``enter itself''.  The other routes
are disambiguation routes, automatically generated by the algorithm
described in Section~\ref{sec:disambiguating}.  These entries are copies
of those present in the main routing table, and prevent locally generated
packets destined to local subnets from leaving through the native default
route.

\begin{figure}
{\scriptsize
\begin{verbatim}
# ip rule show
0:	from all lookup local
101:	from 192.168.4.0/24 lookup 11
32766:	from all lookup main
32767:	from all lookup default
# ip route show
default via 172.23.47.254 dev eth1 proto static
172.23.32.0/20 dev eth1 proto kernel src 172.23.36.138
192.168.4.20 via 192.168.4.20 dev tun-ariane proto 42 onlink
192.168.4.30 via 192.168.4.30 dev wlan1 proto 42 onlink
[...]
# ip route show table 11
default via 192.168.4.20 dev tun-ariane proto 42 onlink
192.168.4.20 via 192.168.4.20 dev tun-ariane proto 42 onlink
192.168.4.30 via 192.168.4.30 dev wlan1 proto 42 onlink
[...]
\end{verbatim}
}
\caption{IPv4 routing table on a router using a VPN}
\label{fig:routing-table}
\end{figure}

\subsection{Multipath TCP} \label{sec:mptcp}

Multipath TCP~\cite{mptcp2012} is an extension to TCP which multiplexes a single
application-layer flow over multiple network layer sub-flows, and attempts to
use as many distinct routes as possible, and to either carry traffic over the
most efficient one or to perform load balancing.  An obvious application is
a mobile node (a telephone) with permanent connectivity to a cellular network
and intermittent WiFi connectivity: MPTCP is able to use the cellular link when
WiFi is not available, and switch to WiFi when available without dropping
already established connections.

Multipath TCP and source-specific routing turn out to be a surprisingly good
match.  MPTCP is able to use all of the addresses of the local host, and to
dynamically probe the reliability and performance of packets sourced from each.

\begin{figure}
  \includegraphics[width=\linewidth]{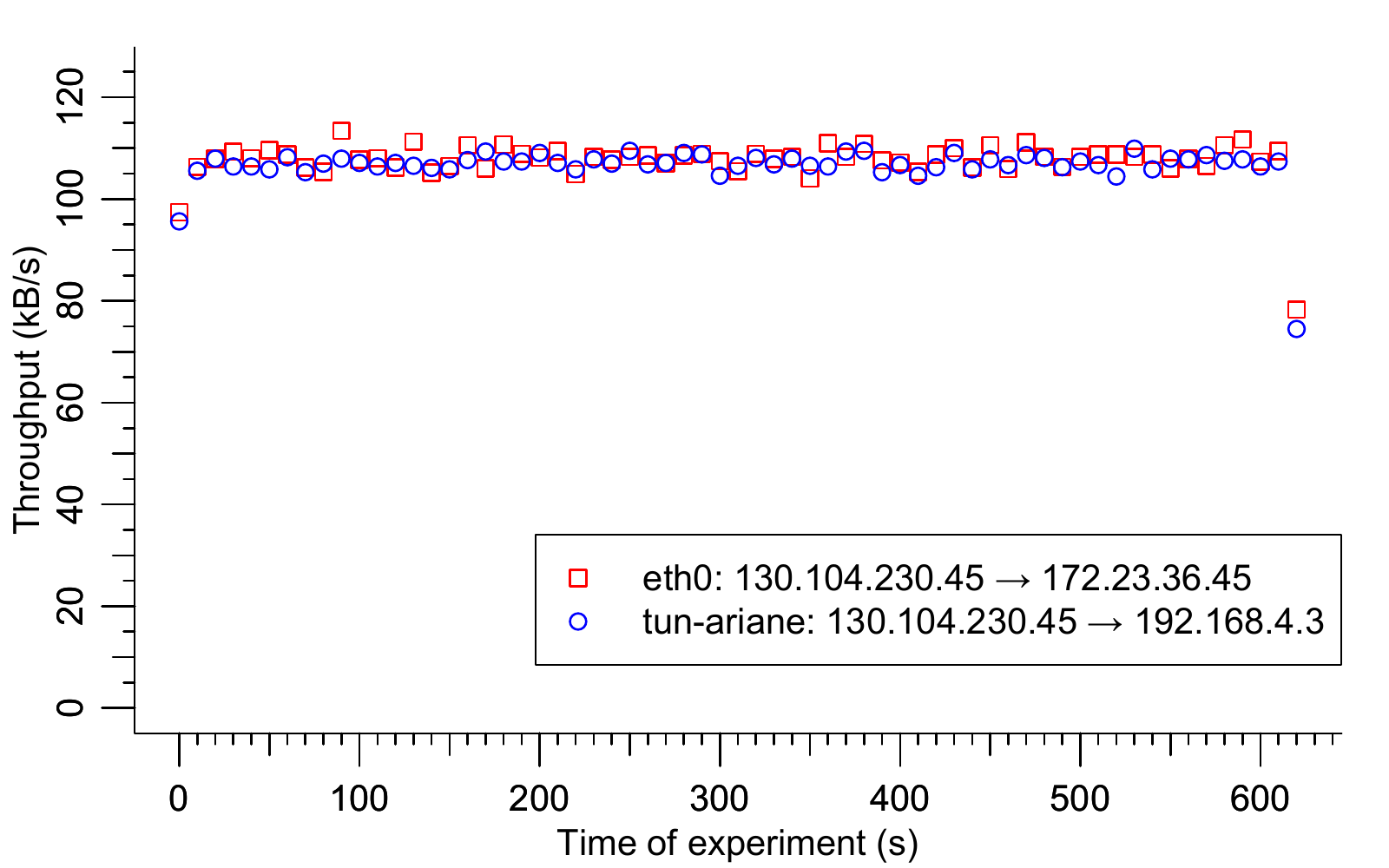}
\caption{Download using MPTCP and traffic control}
\label{fig:mptcp-stable.pdf}
\end{figure}

We have performed two tests that both consist in downloading a 110\,MB
file over MPTCP from the MPTCP website.  In the first test
(Figure~\ref{fig:mptcp-stable.pdf}), a desktop computer is directly
connected to the source-specifically routed wired network, and is configured
with two IPv4 addresses.  The Linux \verb|tc| subsystem is used to limit
each of the addresses to 100\,kB/s traffic; MPTCP is able to reliably
download at 200\,kB/s.

\begin{figure}
  \includegraphics[width=\linewidth]{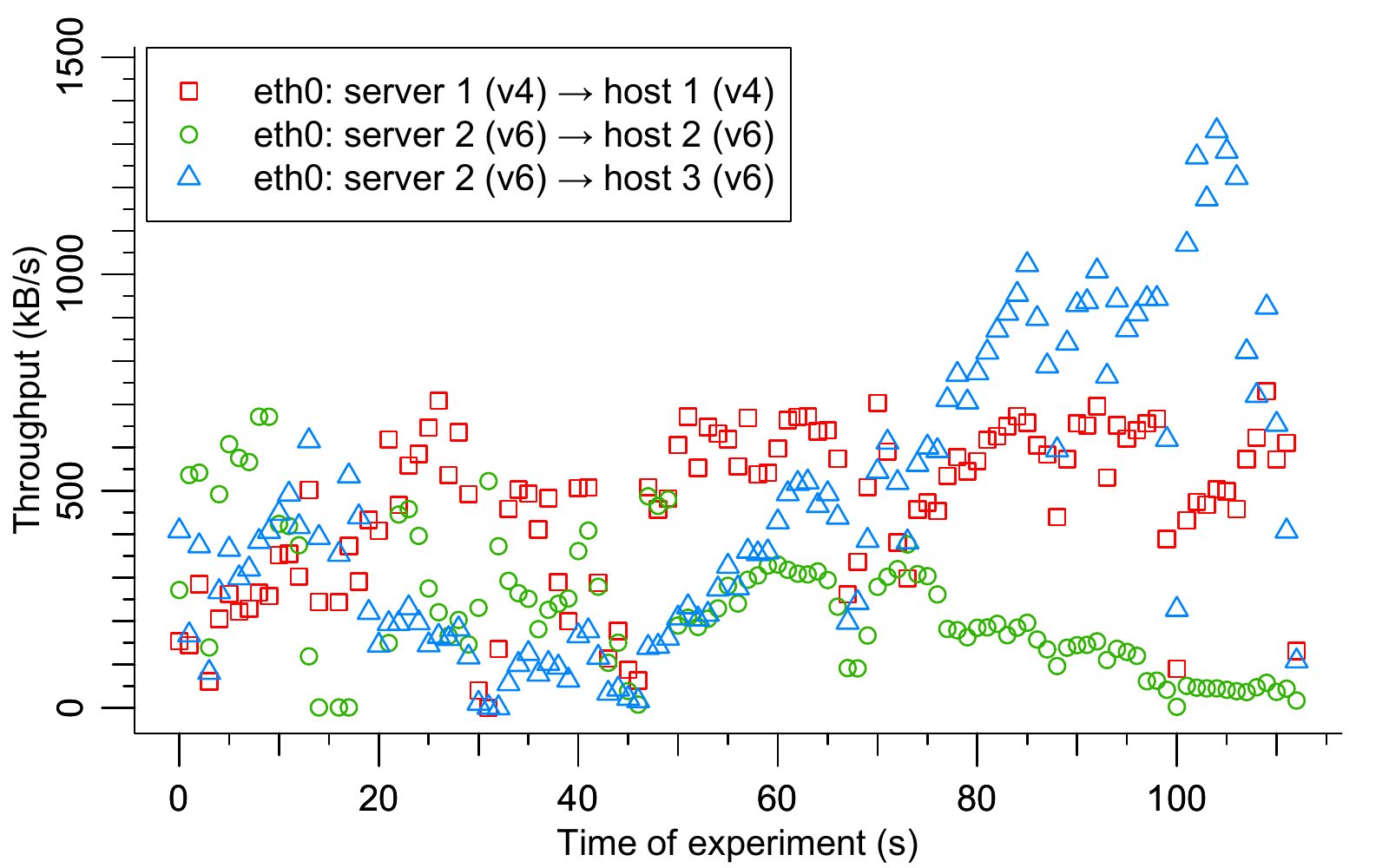}
\caption{Download using MPTCP}
\label{fig:mptcp.pdf}
\end{figure}

In the second test (Figure~\ref{fig:mptcp.pdf}), a laptop's WiFi interface is
configured with three addresses (one IPv4 and two IPv6).  MPTCP multiplexes the
traffic across the three routes, and balances their throughput dynamically.

\section{Conclusion and further work}

Source-specific routing is a modest extension to next-hop routing that
keeps the forwarding decisions firmly within control of the routers while
allowing end hosts a moderate and clearly defined amount of control over
the choice of routes.  Since source-specific routing can cause ambiguous
routing tables, we have defined the behaviour that we believe
source-specific routers should have, and shown how combining different
behaviours in the same network can cause persistent routing loops.
Similar care must be taken when combining non-specific with
source-specific routers in the same network.  We have proposed two ways
to implement source-specific routing, and obtained experimental results
that show that source-specific routing can be usefully exploited by the
transport layer protocol MPTCP.  Our implementation is of production
quality, and has seen a modest amount of deployment, notably as a testbed
for the ideas of the IETF Homenet working group.

While we enjoy working with distance-vector protocols, much of the networking
community appears to have converged on using the OSPF protocol for internal
routing.  OSPF is a rich and complex protocol, and while many of our techniques
should apply without difficulty to it, actually implementing a full
source-specific variant of OSPF without sacrificing any of its flexibility
remains a challenging endeavour.

It was a pleasant surprise to discover that unmodified MPTCP can use
source-specific routes without any manual configuration.  However, we
claim that source-specific routing can also be exploited at the
application layer, and we are currently working on an extension to the
\emph{Mosh} \cite{winstein2012mosh} UDP-based remote shell that is able to
dynamically balance over multiple source-specific routes.

Finally, we have only considered the applicability of source-specific
routing to edge networks, which tend to carry only a moderate number of
distinct routes.  However, there is nothing in principle that would
prevent source-specific routing from being applicable to BGP and to core
networks, where it could perhaps be used for some forms of multihoming and
traffic engineering without the routing table growth due to classical
multihoming.  Extending our results to core networks, with their large
routing tables, will require careful analysis of the complexity of our
techniques, and a carefully optimised implementation.

\section*{Code availability}
The source-specific version of Babel is available from\break{}
\url{https://github.com/jech/babeld}.

\section*{Acknowledgements}

We are grateful to Beno\^it Valiron for his help with the presentation of
the disambiguation algorithm.

\end{document}